\documentclass[11pt]{article}
\usepackage[margin=1in]{geometry}
\usepackage[T1]{fontenc}
\usepackage{lmodern}
\usepackage{amsmath,amssymb}
\usepackage{graphicx}
\usepackage{booktabs}
\usepackage{natbib}
\setcitestyle{authoryear,round,semicolon,aysep={},yysep={,}}
\usepackage[hidelinks]{hyperref}
\usepackage{xurl}
\usepackage{titlesec}
\titleformat{\section}{\normalfont\bfseries\large}{\thesection}{0.6em}{}
\titleformat{\subsection}{\normalfont\bfseries}{\thesubsection}{0.6em}{}
\titlespacing*{\section}{0pt}{1.2ex plus .5ex}{0.6ex}
\titlespacing*{\subsection}{0pt}{1ex plus .5ex}{0.4ex}
\begin{document}

\begin{center}

{\LARGE\bfseries ADZE 2.0: rarefaction estimates of allelic richness and private allelic richness for whole-genome data\par}
\vspace{1.5ex}
{\large Zachary A. Szpiech$^{1,*}$}\\[0.6ex]
{\small $^{1}$Department of Biology, The Pennsylvania State University, University Park, PA 16801 USA\\
$^{*}$zps5164@psu.edu; ORCID: 0000-0001-6372-8224}
\end{center}
\vspace{1ex}
\noindent\textbf{Non-specialist summary}\\
Counting the different versions of a gene, or alleles, found in a population is a simple way to measure its genetic diversity. Alleles found in only one population, or shared only by a few populations, also carry information about how populations are related and about their history. However, the more individuals are sampled, the more alleles are found, so populations sampled at different sizes cannot be compared directly. The program ADZE corrects for this by estimating how many alleles each population would show if all had been sampled at the same size. Here we describe a new version of ADZE that can analyze whole-genome sequencing data from thousands of individuals, and that can also measure diversity region by region along the genome.
\\\\
\noindent\textbf{Abstract}\\
Allelic richness and private allelic richness describe how alleles are distributed across populations, but both depend strongly on sample size. Rarefaction makes them comparable by computing the expected number of alleles in subsamples of a common size, and ADZE implemented this approach and an extension to alleles shared only by a set of populations. However, ADZE was developed for microsatellite data and cannot read sequencing data in variant call format (VCF). Here we present ADZE 2.0, which reads VCF, computes each statistic genome-wide or in genomic windows, and can treat blocks of phased SNPs as haplotype alleles. We also implement improved memory handling and multi-threading support. ADZE 2.0 reproduces the ADZE 1.0 estimates for human microsatellite data exactly. On chromosome 22 of the high-coverage 1000 Genomes Project data, private allelic richness for all 325 population pairs took 4.4 minutes on 32 threads. ADZE 2.0 is available at \url{https://github.com/szpiech/ADZE}.\\

\section{Introduction}
Counts of alleles are among the simplest summaries of genetic variation, and comparing them across populations is common both in conservation genetics \citep{petit1998,leberg2002,foulley2006,caballero2010} and in studies of population history \citep{schroeder2007,jakobsson2008,szpiech2008,szpiech2011}. However, the number of alleles seen in a sample grows with the number of gene copies sampled. A population that is sampled deeply will tend to identify more alleles compared to a population that is sampled sparsely. Rarefaction corrects for unequal sampling by calculating how many alleles would be seen if every population had been sampled at the same size $g$, where $g$ is no larger than the smallest sample. In ecology, rarefaction is used to compare species counts between samples of different sizes \citep{hurlbert1971}, and \citet{petit1998} applied it to allelic richness. \citet{kalinowski2004} then defined a rarefied count of private alleles, and both measures are computed by the program HP-RARE \citep{kalinowski2005}.

Private alleles carry information about how populations are related \citep{Slatkin1985,Smith2026}. However, among closely related populations an allele is rarely confined to one population \citep{kalinowski2004}, and it is more often shared by two or three populations and absent from the rest. \citet{szpiech2008} therefore extended rarefied private allelic richness from single populations to sets of populations, counting the alleles that occur in every member of a set and in no population outside it. For a set of one population, this is the private allelic richness of \citet{kalinowski2004}. When computed for the entire powerset of populations, this provides a complete partitioning of alleles shared across populations \citep{jakobsson2008,szpiech2008}. Allelic richness, private allelic richness and its extension to sets of populations were implemented in the program ADZE (Allelic Diversity AnalyZEr) \citep{szpiech2008}.

ADZE 1.0 was developed for microsatellite data in the format used by \textsc{structure} \citep{pritchard2000}. The dimensions of the data file had to be written in a parameter file, every genotype was held in memory, and the time for locus filtering grew with the square of the number of loci. Modern whole-genome sequencing datasets, such as the high-coverage 1000 Genomes Project data \citep{byrskabishop2022}, now contain millions of variant sites for thousands of individuals and are distributed in the variant call format (VCF) \citep{danecek2011}, which ADZE 1.0 cannot read and would struggle to load into memory.

Here we present an update to ADZE with improved memory usage, computational efficiency, and usability. ADZE 2.0 reads VCF files directly, can compute each statistic genome-wide or in windows along the genome, and can analyze haplotypes by grouping consecutive phased SNPs into alleles. Parameter files written for ADZE 1.0 still work without changes. We describe the estimators, the new features, a re-analysis of the human microsatellite data of \citet{szpiech2008}, and the run time and memory use of ADZE 2.0 on chromosome 22 of the high-coverage 1000 Genomes Project data.

\section{Methods}
We follow the notation of \citet{szpiech2008}. At a given locus, let $I$ be the number of allele types observed, and let $N_{ij}$ be the count of allele type $i$ among the gene copies sampled from population $j$, so that population $j$ has $N_j=\sum_{i=1}^{I}N_{ij}$ sampled gene copies at the locus. If $g$ of these gene copies are sub-sampled at random without replacement, allele type $i$ is absent from the sub-sample with probability
\begin{equation}
  Q_{ijg}=\frac{\binom{N_j-N_{ij}}{g}}{\binom{N_j}{g}},
  \label{eq:Q}
\end{equation}
and present with probability $P_{ijg}=1-Q_{ijg}$. Summing $P_{ijg}$ over allele types gives the rarefied allelic richness of population $j$, which is the expected number of allele types in the sub-sample \citep{hurlbert1971,petit1998,kalinowski2004},
\begin{equation}
  \hat{\alpha}_g^{(j)}=\sum_{i=1}^{I}P_{ijg}.
  \label{eq:alpha}
\end{equation}
An allele is private to population $j$ at size $g$ if it appears in $j$ and not in the other $J-1$ populations. The rarefied private allelic richness is the expected number of such alleles \citep{kalinowski2004},
\begin{equation}
  \hat{\pi}_g^{(j)}=\sum_{i=1}^{I}\left[P_{ijg}\left(\prod_{\substack{j'=1\\ j'\neq j}}^{J}Q_{ij'g}\right)\right],
  \label{eq:pi}
\end{equation}
To extend this to sets of populations, write $S=\{1,2,\ldots,J\}$ for the full set of populations, and write $C_{km}$, with $m=1,\ldots,\binom{J}{k}$, for its subsets of size $k$, which together make up $C_k$. An allele is private to $C_{km}$ if it appears in every population in $C_{km}$ and in none of the populations in $S\setminus C_{km}$. The expected number of these alleles is \citep{szpiech2008}
\begin{equation}
  \hat{\pi}_{gk}^{(m)}=\sum_{i=1}^{I}\left[\left(\prod_{j\in C_{km}}P_{ijg}\right)\left(\prod_{j'\in S\setminus C_{km}}Q_{ij'g}\right)\right],
  \label{eq:pigk}
\end{equation}
Setting $k=1$ recovers Equation~(\ref{eq:pi}), and setting $k=J-1$ counts the alleles that are absent from one population and present in all of the others \citep{szpiech2008}. Note that every quantity above is an exact expectation, so no random subsampling is done and the results do not change from run to run.

Each statistic is computed at every locus, and ADZE reports the mean across the $L$ loci used, $\bar{x}=L^{-1}\sum_{\ell=1}^{L}x_\ell$, together with the variance across loci, $s^2=(L-1)^{-1}\sum_{\ell=1}^{L}(x_\ell-\bar{x})^2$, and the standard error of the mean, $s/\sqrt{L}$, at each $g$ and for each population or combination. A statistic is only defined at a given $g$ if it is defined at that $g$ at every locus used. For private allelic richness, this means that $g$ can be no larger than the smallest $N_j$ across populations at each locus. By default, the largest standardized sample size, \texttt{MAX\_G}, is therefore the smallest $N_j$ over all populations and all loci that remain after missing-data filtering.

\section{Features of ADZE 2.0}

\subsection{Usability}
Every parameter can be given on the command line, and only the data file is required. Parameter files written for ADZE 1.0 work without changes, with the same keywords and the same short flags, and an option given on the command line takes precedence over the parameter file. ADZE 1.0 required the number of loci, the number of data rows and the number of label columns to be given, and ADZE 2.0 now measures them as it reads the file. The grouping column is the last label column by default, and \texttt{MAX\_G} is by default the largest value the data support. Any of these values can still be given, and a warning is printed if a given value differs from the measured one. \texttt{-{}-dry-run} reports the measured dimensions, the sample size of each grouping, the largest feasible \texttt{MAX\_G} and the number of combinations an analysis would evaluate, and then stops without computing any statistics.

Output files are named from \texttt{-{}-out-prefix}, and \texttt{-{}-stat} selects which statistics are computed. \texttt{-{}-at-g} reports a single $g$, either a given value or \texttt{max}, so that the value of \texttt{MAX\_G} does not need to be known in advance. \texttt{-{}-pops} restricts an analysis to a list of groupings, and \texttt{-{}-exclude-pops} excludes groupings from the analysis. Note that both options change $J$, and so they also change which alleles count as private. Combinations of populations can be requested by size with \texttt{-{}-tuples-k} or listed in a file with \texttt{-{}-tuples}. The results files are tab-separated with a header, and a statistic that is undefined at a given $g$ is written as \texttt{NA}. \texttt{-{}-legacy} writes the ADZE 1.0 layout of these files instead.

Missing data are handled as in ADZE 1.0, but the default and the reporting differ. \texttt{-{}-tolerance} $x$ drops any locus at which some grouping has more than a fraction $x$ of its gene copies missing, and the dropped loci are listed in a separate file. Because a statistic is only defined at $g$ if it is defined at every locus, a single locus at which one grouping has high missing data limits the whole analysis. The default \texttt{-{}-tolerance} is now $x=0.1$, rather than the value of 1 used by ADZE 1.0, which kept every locus. When a retained locus makes a grouping unusable, ADZE reports the grouping by name with the number of loci involved, and, if filtering removes every locus, ADZE stops with an error.

\subsection{VCF input}
ADZE 2.0 reads VCF files, uncompressed or compressed with \texttt{gzip} or \texttt{bgzip}. Only the \texttt{GT} field is read, so any filtering of sites or genotypes must be done before the file is given to ADZE. A sample contributes as many gene copies as its \texttt{GT} field holds, so haploid and diploid records can be in the same file, and an uncalled allele counts as missing under \texttt{-{}-tolerance} in the same way as for \textsc{structure} input.

Because a VCF does not contain population information, \texttt{-{}-samples} is required with VCF input. It takes a file with two columns, the sample name and the grouping name. A sample in the VCF that is not in the sample file is skipped, and a sample in the file that is not in the VCF is ignored, with a warning that gives a count for each case. The same VCF can therefore be analyzed for a subset of samples, or at a different level of grouping, by using a different file.

\subsection{Haplotype alleles}
\label{sec:haplotypes}
With phased VCF input, consecutive SNPs can be combined into blocks, and each block is a locus whose alleles are haplotypes of the SNP alleles. \texttt{-{}-hap-snps} $N$ makes each block $N$ consecutive SNPs, and \texttt{-{}-hap-bp} $N$ makes each block a region of $N$ basepairs starting at position 1. Blocks do not overlap and do not span chromosomes, and \texttt{-{}-min-hap-snps} $N$ drops blocks that contain fewer than $N$ SNPs. When using haplotypes, an additional output file records, for every block, the number of SNPs, the number of distinct haplotypes and the number of haploid samples in each grouping that carry a complete haplotype. Haplotype blocks can be used with windows (Section~\ref{sec:windows}), where each block is assigned to the window that contains its first SNP.

\subsection{Windows along the genome}
\label{sec:windows}
Each statistic can also be reported in windows along the genome, in addition to being reported genome-wide. \texttt{-{}-window-bp} $N$ gives windows of $N$ basepairs starting at position 1, so the same region gives the same windows in every run. \texttt{-{}-window-loci} $N$ gives windows of $N$ loci that remain after filtering. In both cases, the step between windows is set with \texttt{-{}-step-bp} or \texttt{-{}-step-loci}, and it is equal to the window size by default. Positions are taken from the VCF, or from a separate file given with \texttt{-{}-loci-map} for \textsc{structure} input.

\texttt{MAX\_G} is determined once from all loci that remain after filtering, and the same value is used for every window. A window in a region with larger sample sizes is therefore not reported at a larger $g$ than the rest of the genome, and all windows along a scan can be compared at the same $g$. Because the variance and standard error are computed across the loci in a window, they are noisy when a window contains few loci, and \texttt{-{}-min-window-loci} sets the smallest number of loci a window must contain to be reported. Each statistic writes its windows to a separate file with \texttt{\_windows} added to its name.

\subsection{Memory and threads}
ADZE 2.0 does not hold the genotypes in memory, but, instead, it reads the data twice. The first pass determines \texttt{MAX\_G}, the windows, and which loci to filter. The second pass computes every statistic one locus at a time and keeps a running mean and variance, so memory use does not explode with increasing sample size. The option \texttt{-{}-threads} $N$ divides the loci among $N$ threads from the C++ standard library. 

\section{Results}

\subsection{Re-analysis of the 2008 HGDP data}
To check that ADZE 2.0 computes the same quantities as ADZE 1.0, we re-analyzed the human microsatellite data of \citet{szpiech2008}. These data are the 783 microsatellite loci of \citet{rosenberg2005} genotyped in the 952 individuals of the HGDP-CEPH panel with no known first- or second-degree relatives \citep{rosenberg2006}, which we grouped into the same five geographic regions (Africa, Eurasia, East Asia, Oceania and America). As in \citet{szpiech2008}, we kept the loci with at most 15\% missing data in each region (\texttt{-{}-tolerance 0.15}), which left 721 loci, and we computed $\hat{\alpha}_g^{(j)}$, $\hat{\pi}_g^{(j)}$ and $\hat{\pi}_{gk}^{(m)}$ for $k=2,\ldots,5$ at every $g$ up to the largest value these loci allow ($g=48$). Both versions were run on the same \textsc{structure} file with the same parameter file.

\begin{figure}[t]
\centering
\IfFileExists{figures/hgdp_reproduction.pdf}{%
  \includegraphics[width=\textwidth]{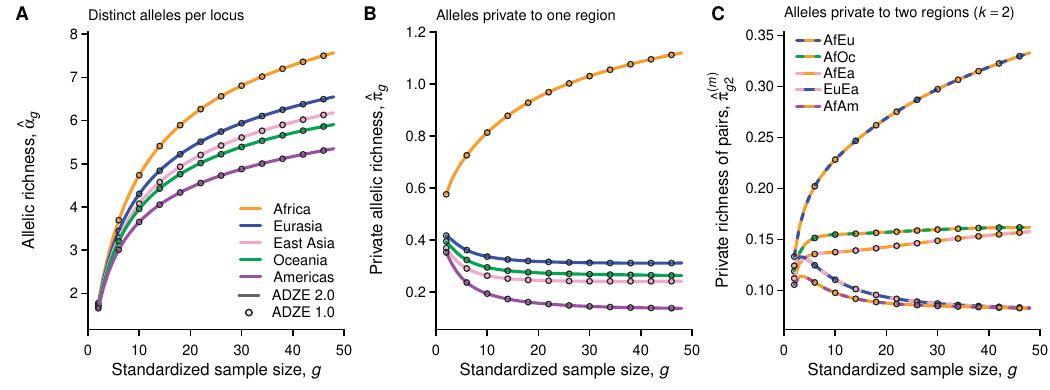}%
}{%
  \fbox{\parbox[c][1.8in][c]{0.92\columnwidth}{\centering\small Figure file \texttt{figures/hgdp\_reproduction.pdf} not yet produced.}}%
}
\caption{Estimates from ADZE 1.0 (open circles) and ADZE 2.0 (lines) for 721 microsatellite loci in 952 HGDP-CEPH individuals grouped into five regions. (A) Allelic richness $\hat{\alpha}_g$. (B) Private allelic richness $\hat{\pi}_g$. (C) Private allelic richness for pairs of regions, $\hat{\pi}_{g2}^{(m)}$, for the five pairs with the largest values at $g=40$. Each pair is drawn in the colors of its two regions. Version 1.0 values are shown at every fourth $g$.}
\label{fig:hgdp}
\end{figure}

ADZE 1.0 and ADZE 2.0 removed the same 62 loci, and every one of the 5076 values that both versions report (means, variances and standard errors) is identical (Fig.~\ref{fig:hgdp}). The estimates also reproduce those published by \citet{szpiech2008}. For example, all 124 values in their Table 1 at $g=10$, 20, 30 and 40 match to the two decimal places printed (Table~\ref{tab:hgdp}), including the larger share of private alleles in the combination of Africa and Oceania than in the combination of Africa and East Asia at $g=40$ (1.80\% and 1.71\%). On this data set, ADZE 2.0 took 0.17 s and 7.6 MB of memory on one thread, and ADZE 1.0 took 3.01 s and 57 MB.

\begin{table}[t]
\centering
\caption{Percentage of distinct alleles private to each combination of the five geographic regions in the 952 HGDP-CEPH individuals (721 loci), as reported in Table~1 of \citet{szpiech2008} and as computed with ADZE 2.0. Af, Africa; Eu, Eurasia; Ea, East Asia; Oc, Oceania; Am, America. For each combination, the rarefied value at standardized sample size $g$ is $\hat{\pi}_{gk}^{(m)}$ (for single regions, $\hat{\pi}_g^{(j)}$) expressed as a percentage of the sum over all 31 combinations at that $g$. Uncorrected values are the percentages of the 8516 distinct alleles in the sample, as reported by \citet{szpiech2008}.}
\label{tab:hgdp}
\small
\setlength{\tabcolsep}{4pt}
\begin{tabular}{l r rr rr rr rr}
\toprule
 & & \multicolumn{2}{c}{$g=10$} & \multicolumn{2}{c}{$g=20$} & \multicolumn{2}{c}{$g=30$} & \multicolumn{2}{c}{$g=40$}\\
\cmidrule(lr){3-4}\cmidrule(lr){5-6}\cmidrule(lr){7-8}\cmidrule(lr){9-10}
Geographic distribution & Uncorrected & 2008 & v2.0 & 2008 & v2.0 & 2008 & v2.0 & 2008 & v2.0\\
\midrule
AfEuEaOcAm & 41.67 & 22.95 & 22.95 & 33.65 & 33.65 & 38.36 & 38.36 & 41.00 & 41.00 \\
\addlinespace
AfEuEaOc & 5.70 & 4.51 & 4.51 & 5.18 & 5.18 & 5.35 & 5.35 & 5.37 & 5.37 \\
AfEuEaAm & 7.40 & 3.92 & 3.92 & 3.84 & 3.84 & 3.66 & 3.66 & 3.53 & 3.53 \\
AfEuOcAm & 0.18 & 2.67 & 2.67 & 2.22 & 2.22 & 1.90 & 1.90 & 1.69 & 1.69 \\
AfEaOcAm & 0.04 & 2.60 & 2.60 & 1.93 & 1.93 & 1.55 & 1.55 & 1.31 & 1.31 \\
EuEaOcAm & 0.54 & 4.11 & 4.11 & 2.99 & 2.99 & 2.34 & 2.34 & 1.95 & 1.95 \\
\addlinespace
AfEuEa & 8.14 & 2.27 & 2.27 & 2.31 & 2.31 & 2.33 & 2.33 & 2.36 & 2.36 \\
AfEuOc & 0.59 & 2.25 & 2.25 & 2.15 & 2.15 & 2.04 & 2.04 & 1.97 & 1.97 \\
AfEuAm & 0.76 & 1.63 & 1.63 & 1.34 & 1.34 & 1.20 & 1.20 & 1.13 & 1.13 \\
AfEaOc & 0.08 & 1.65 & 1.65 & 1.43 & 1.43 & 1.30 & 1.30 & 1.22 & 1.22 \\
AfEaAm & 0.04 & 1.29 & 1.29 & 0.96 & 0.96 & 0.83 & 0.83 & 0.76 & 0.76 \\
AfOcAm & 0.02 & 1.04 & 1.04 & 0.73 & 0.73 & 0.60 & 0.60 & 0.54 & 0.54 \\
EuEaOc & 0.88 & 1.66 & 1.66 & 1.19 & 1.19 & 0.97 & 0.97 & 0.86 & 0.86 \\
EuEaAm & 0.96 & 1.63 & 1.63 & 1.07 & 1.07 & 0.84 & 0.84 & 0.73 & 0.73 \\
EuOcAm & 0.04 & 1.08 & 1.08 & 0.63 & 0.63 & 0.46 & 0.46 & 0.38 & 0.38 \\
EaOcAm & 0.02 & 1.29 & 1.29 & 0.79 & 0.79 & 0.59 & 0.59 & 0.50 & 0.50 \\
\addlinespace
AfEu & 7.15 & 3.23 & 3.23 & 3.34 & 3.34 & 3.46 & 3.46 & 3.55 & 3.55 \\
AfEa & 1.17 & 1.95 & 1.95 & 1.77 & 1.77 & 1.73 & 1.73 & 1.71 & 1.71 \\
AfOc & 0.26 & 2.19 & 2.19 & 1.95 & 1.95 & 1.86 & 1.86 & 1.80 & 1.80 \\
AfAm & 0.18 & 1.38 & 1.38 & 1.09 & 1.09 & 0.99 & 0.99 & 0.93 & 0.93 \\
EuEa & 3.10 & 1.56 & 1.56 & 1.16 & 1.16 & 1.01 & 1.01 & 0.94 & 0.94 \\
EuOc & 0.38 & 1.48 & 1.48 & 1.04 & 1.04 & 0.88 & 0.88 & 0.79 & 0.79 \\
EuAm & 0.28 & 1.19 & 1.19 & 0.75 & 0.75 & 0.60 & 0.60 & 0.52 & 0.52 \\
EaOc & 0.15 & 1.38 & 1.38 & 0.99 & 0.99 & 0.87 & 0.87 & 0.81 & 0.81 \\
EaAm & 0.22 & 1.22 & 1.22 & 0.79 & 0.79 & 0.64 & 0.64 & 0.55 & 0.55 \\
OcAm & 0.01 & 0.97 & 0.97 & 0.60 & 0.60 & 0.46 & 0.46 & 0.39 & 0.39 \\
\addlinespace
Af & 7.32 & 11.51 & 11.51 & 11.80 & 11.80 & 11.98 & 11.98 & 12.08 & 12.08 \\
Eu & 8.20 & 4.76 & 4.76 & 3.94 & 3.94 & 3.63 & 3.63 & 3.45 & 3.45 \\
Ea & 3.36 & 3.71 & 3.71 & 3.03 & 3.03 & 2.80 & 2.80 & 2.67 & 2.67 \\
Oc & 0.58 & 4.16 & 4.16 & 3.40 & 3.40 & 3.11 & 3.11 & 2.94 & 2.94 \\
Am & 0.62 & 2.73 & 2.73 & 1.94 & 1.94 & 1.67 & 1.67 & 1.54 & 1.54 \\
\bottomrule
\end{tabular}
\end{table}

\subsection{Runtime}
To measure the run time of ADZE 2.0 on whole-genome sequencing data, we analyzed the phased VCF file for chromosome 22 of the high-coverage 1000 Genomes Project data \citep{byrskabishop2022}. This file contains 1\,070\,401 records for 3202 individuals, and we used \texttt{-{}-samples} to keep the 2504 unrelated individuals grouped into their 26 populations. We timed four analyses: $\hat{\alpha}_g^{(j)}$ and $\hat{\pi}_g^{(j)}$ for the whole chromosome, the same in 1-Mb windows (39 windows), and each of these with $\hat{\pi}_{gk}^{(m)}$ for all 325 pairs of populations ($k=2$). We ran each analysis with every variant site as a locus and again with each block of 10 consecutive sites as a locus (i.e., 10-variant haplotypes, \texttt{-{}-hap-snps 10} and \texttt{-{}-min-hap-snps 10}, 107\,040 blocks). All other options were left at their defaults, so \texttt{MAX\_G} was 122, the number of gene copies in the smallest population, and no locus was removed by \texttt{-{}-tolerance 0.1}. Each analysis was run once at 1, 2, 4, 8, 16 and 32 threads on a machine with one 18-core Intel Xeon W-2295 processor (36 hardware threads) and 62 GB of memory. Elapsed time and peak memory were measured with \texttt{/usr/bin/time}.

On one thread, allelic richness and private allelic richness for the whole chromosome took 321 s, and adding all 325 pairs of populations raised this to 2052 s (Table~\ref{tab:runtime}). Reporting the same statistics in windows added between 16\% and 34\% to the time on one thread. With 32 threads, the analyses that include pairs of populations ran 5.8 to 7.7 times faster than on one thread, and the analysis of all pairs at every variant site took 266 s. The speed-up was smaller for the analyses without pairs (2.7 to 4.5 times on 32 threads), likely because they do less computation per locus, so that more of their time is spent reading the file, which is done on a single thread. Peak memory did not exceed 507 MB in any run. It increased with the number of threads, and on one thread it was 28 to 71 MB for the haplotype loci.

\begin{table}[t]
\centering
\caption{Elapsed time of ADZE 2.0 on chromosome 22 of the high-coverage 1000 Genomes Project data (2504 individuals in 26 populations, \texttt{MAX\_G} = 122) by number of threads. Times are in seconds from one run of each configuration. Windows are 1 Mb and do not overlap (39 windows). Speed-up is the time on one thread divided by the time on 32 threads, and the last column gives the range of peak memory over the six thread counts.}
\label{tab:runtime}
\small
\begin{tabular}{l rrrrrr r r}
\toprule
 & \multicolumn{6}{c}{Elapsed time (s) by number of threads} & & Peak\\
\cmidrule(lr){2-7}
Statistics & 1 & 2 & 4 & 8 & 16 & 32 & Speed-up & memory (MB)\\
\midrule
\multicolumn{9}{l}{\textit{Each variant site a locus (1\,070\,401 loci)}}\\
$\hat{\alpha}$, $\hat{\pi}$ & 321 & 250 & 164 & 106 & 104 & 79 & 4.0 & 201--471\\
$\hat{\alpha}$, $\hat{\pi}$, windows & 431 & 273 & 185 & 125 & 121 & 96 & 4.5 & 215--459\\
$\hat{\alpha}$, $\hat{\pi}$, 325 pairs & 2\,052 & 1\,196 & 719 & 417 & 318 & 266 & 7.7 & 220--505\\
$\hat{\alpha}$, $\hat{\pi}$, 325 pairs, windows & 2\,386 & 1\,339 & 869 & 555 & 471 & 410 & 5.8 & 247--507\\
\multicolumn{9}{l}{\textit{Each block of 10 sites a locus (107\,040 loci)}}\\
$\hat{\alpha}$, $\hat{\pi}$ & 360 & 233 & 177 & 147 & 133 & 127 & 2.8 & 28--374\\
$\hat{\alpha}$, $\hat{\pi}$, windows & 346 & 234 & 180 & 148 & 136 & 129 & 2.7 & 30--369\\
$\hat{\alpha}$, $\hat{\pi}$, 325 pairs & 1\,285 & 717 & 430 & 277 & 213 & 199 & 6.4 & 29--225\\
$\hat{\alpha}$, $\hat{\pi}$, 325 pairs, windows & 1\,296 & 737 & 447 & 300 & 229 & 216 & 6.0 & 71--262\\
\bottomrule
\end{tabular}
\end{table}

\section{Implementation and availability}
ADZE 2.0 is written in C++ and requires only a C++17 compiler. If zlib is available, it is used to read compressed input, and without it the program builds and runs and refuses a compressed file with an error message. It can be built with CMake or with \texttt{make}. The source code, the manual and the example data are available at \url{https://github.com/szpiech/ADZE}. Stand-alone binaries for Linux, macOS and Windows are also provided. License: GPLv3.

\section*{Acknowledgments}
We would like to thank T. Quinn Smith for testing ADZE 2.0.

\section*{Funding}
This work was supported by the National Institute of General Medical Sciences of the National Institutes of Health under Award Number R35GM146926.

\bibliographystyle{pbmt}
\bibliography{adze2_appnote}

\end{document}